# Energetics of Cs in Σ3 grain boundary of 3C-SiC

Raghani Pushpa[1*], Balaji Ramanujam[1], Tram Bui[2], and Megan Frary[2]

[1]*Department of Physics, Boise State University, 1910 University Dr., Boise, Idaho, USA*

[2]*Material Science and Engineering, Boise State University, 1910 University Dr., Boise, Idaho, USA*

**Abstract**

Energetics of Cs defects at Σ3 grain boundaries of 3C-SiC has been studied using density functional theory to understand the role of the grain boundaries in Cs diffusion and its eventual release from the tristructural isotropic fuel particles (TRISO). Cs is shown to be much more stable at the Σ3 grain boundary than in bulk of SiC with a significant decrease (7 – 17 eV) in the formation energies at grain boundaries than in bulk. It is found to have even lower formation energies than those of Ag at the Σ3 grain boundaries, while this trend was opposite in the bulk SiC as demonstrated previously from similar density functional theory calculations. Based on these results, a possible route to control Cs release from SiC layer via grain-boundary-engineering is suggested.

*Corresponding author email: pushparaghani@boisestate.edu

1. **Introduction**

The cubic polytype silicon carbide (3C-SiC) is a crucial coating material in the production of tristructural isotropic (TRISO) fuel particles used in the very high temperature reactors (VHTR). The 3C-SiC coating of 35 μm [1] in TRISO acts as the structural component for providing mechanical support to the soft coatings of inner and outer pyrolytic carbon and as the barrier for controlling the diffusion of fission products (FPs) like Cs, I, Ag, Sr & Kr [2, 3]. Several hypotheses on the diffusion of these fission products through SiC are reported in the literature for I [4-7], Cs [8-13], Ag [6, 11, 14-21], Sr [7], and Kr [3, 22, 23]. Additionally, different diffusion mechanisms/pathways for the FPs that are probable for causing chemical degradation of SiC [24, 25], vapor diffusion through cracks and nanopores [17, 26, 27], and diffusion through grain boundaries (GBs) and (pipe) dislocation [28, 29] are also reported. Despite SiC being the potential barrier for the FPs, there are reports on the diffusion of fission products through the SiC layer [5, 6, 18, 25, 30] depending on the reactor operation conditions.

Among many types of fission products, Cs and Ag have been reported to be the major fission products released during the operational and post-operational conditions of the reaction of TRISO particles. Cs is found to be released under higher temperature (1600 K – 2000 K) operating conditions [8, 11, 26, 31]. This has raised significant concern about the maintenance and safety of the reactor and the environment, especially due to the long radioactive half-life (30 years) of the $^{137}$Cs isotope, high vapor pressure (3 x $10^{-9}$ atm at 302 K) and low boiling point (944K) [32] of the released Cs. The integral release experiments [8, 11, 26, 31] on the Cs release show a wide range of the diffusion coefficients (from $10^{-11}$m$^2$/s to $10^{-2}$ m$^2$/s) along with the change in activation energy (from 1.82 eV/atom to 5.33 eV/atom) at temperatures ranging from 1100 K to 2000 K. A recent report by Shrader *et al*. [9]

suggests that the higher diffusion coefficient with higher activation energy is associated with the Cs movement through the bulk SiC, while the low diffusion coefficient with low activation energy might be associated with Cs movement through the grain boundaries (GBs). They have specifically studied the diffusion of Cs in the bulk SiC using density functional theory. However, Cs diffusion at the grain boundaries of SiC has not been investigated to date and reports in the literature suggest that grain boundaries play a crucial role in the diffusion of fission products through SiC [28, 29].

Among the available low-energy grain boundaries of cubic SiC, the Σ3 grain boundary is known to be a special twin grain boundary. It is stabilized with the presence of 6- or 7-atom rings [20, 33-37]. Of the fission products studied to date, only Ag has been systematically studied in bulk as well as at the Σ3 grain boundary of SiC by Khalil *et al.* [20] using atomistic simulations. In the present work we focus on studying the energetics of Cs at the grain boundary of SiC and qualitatively assess the diffusion of Cs through SiC based on the ease or difficulty of defect formation in the GB compared with that in the bulk. We also compare the results on formation energies of Cs with those of other fission products, in particular Ag, and find that Cs is in fact more stable at the grain boundaries than Ag despite the bigger size of Cs. From this comparison, we also discuss why Cs has very different diffusion coefficients and activation energies at low and high temperatures, i.e., why there is a sudden change in the dominant diffusion mechanism of Cs as a function of temperature.

## 2. Methods

To investigate Cs defects at the Σ3-grain boundary of SiC, Quantum ESPRESSO [38] (open-Source Package for Research in Electronic Structure, Simulation, and Optimization) was used for the DFT calculations in the pseudopotential formalism. The generalized gradient approximation (GGA) is employed to approximate the exchange-correlation potential using the Perdew-Burke-Ernzerhof exchange correlation functional [39]. Ultrasoft pseudopotentials [40] were used to model the interactions between the valence electrons and the nuclei for carbon and silicon atoms, whereas for Cs, a norm-conserving pseudopotential is generated with nonlinear core correction [41] based on the Troullier-Martins recipe [42] was used. The pseudopotentials were generated with the following valance electronic configurations: C as $2s^2\ 2p^2$[43], Si as $3s^2\ 3p^2$[43] and Cs having 0.6 electrons in the valence state as $6s^{0.5}\ 5d^{0.05}\ 6p^{0.05}$[44]. The relaxation of atomic positions has been carried out at a fixed volume with a quasi-Newton algorithm optimization method, having force components converging to less than 0.001 Ry/bohr.

A Monkhorst-Pack grid was used for the Brillouin zone sampling. For calculations in bulk undefected SiC, a *k*-point grid of 10×10×10 was used for a 1×1×1 unit cell having 8 atoms, and the results for lattice constants and bulk moduli were found to be in excellent agreement with those from experiments. For intrinsic or extrinsic defects in the bulk SiC, a 2×2×2 supercell having 64 atoms was used. For investigating defects at the Σ3-grain boundary in SiC, a 2×2×1 slab having 256 atoms was used and the Brillouin zone integrations were performed using a 3x3x3 *k*-point mesh. Along the direction perpendicular to the grain boundary plane, two slabs are separated by a 4.6 Å thick vacuum. The wave functions were expanded using a plane wave cut-off of 600 eV.

The defect formation energies ($E_F$) of neutral Cs in bulk SiC and the Σ3-grain boundary of SiC were calculated using the following equation [21, 45]:

$$E_F = E_{def} - E_{undef} + \sum_i \Delta n_i \mu_i, \qquad (1)$$

where $E_{def}$ & $E_{undef}$ correspond to the calculated total energies of the defected and un-defected systems, respectively. $\Delta n_i$ is the difference in the number of atoms of species 'i' between the undefected (perfect) and the defected super cells, and $\mu_i$ is the chemical potential of species 'i' [45]. The chemical potentials of the Si and C atoms are calculated using the cohesive energies of bulk-Si and graphite respectively, in Si-rich and C-rich conditions. In the present work, however, we focus on the discussion of formation energies in Si-rich conditions only, for the sake of brevity. Formation energies in C-rich conditions can be obtained by adding or subtracting (depending on the defect type) the heat of formation of SiC from the formation energies presented in Si-rich conditions. These cohesive energies (per atom) for the bulk-Si and graphite are found to be -5.40 eV and -9.23 eV, respectively, which compare well with the reported theoretical values of -5.44 eV and -9.20 eV [21]. However, the experimental values of cohesive energies are -4.63 eV and -7.37 eV[46] for bulk-Si and graphite. In addition, the calculated band structure and the density of states of bulk SiC were also found to have a good comparison with those from previous calculations [47, 48].

The electronic band gap for bulk SiC is found to be 1.40 eV from our calculations which is in agreement with other theoretical data of 1.37 eV [47, 48] and 1.53 eV [47, 48] but it is lower than the experimental value of 2.39 eV[49]. This underestimation of the band gap may be due to the underestimation of the electron-electron correlation interaction between electrons in the exchange-correlation functional [50].

## 3. Grain Boundary Structure

The atomic structure of the Σ3-grain boundary (GB) was taken in a bicrystal configuration from Ref. [20]. The Σ3-grain boundary lies in the ($2\bar{1}\bar{1}$) plane, and its tilt axis is aligned with the [$0\bar{1}1$] direction [20]. Figure 1 shows the XZ cross sectional structure of the most stable Σ3-grain boundary with two grains, viz., grain I (GI) and grain II (GII) (separated by a black dashed line in Figure. 1); it consists of 5-, 6-, and 7-membered rings whose interstitial sites are labeled as I1, I3 or I4 and I2, respectively, present at the grain plane and/or at the GI or at the GII as shown in Figure 1. Thickness of the Σ3-GB along z-direction is about 6 Å that contains the 5-, 6-, and 7-membered rings [20]. The Y direction of the Σ3-grain boundary is parallel to the tilt axis $[0\bar{1}1]$, and the X and the Z directions are [111] and $[2\bar{1}\bar{1}]$ as shown in Figure 1. The size of the coincidence site lattice (CSL) unit cell is 7.55 Å in X, 3.08 Å in Y. The size of the whole grain boundary structure is 15.2 Å in X, 12.3 Å in Y, and 18 Å in Z directions. Further, in order to study the Cs as point defects, distinct atomic positions present at the grain plane, GI and GII were considered; they are labeled as C1, C2, C3, C4, Si1, Si2, Si3 and Si4 in Figure 1.

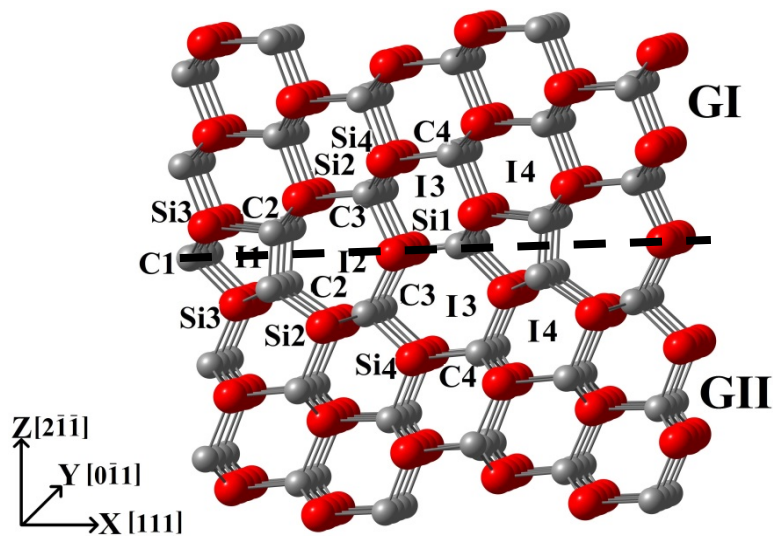

**Figure 1.** The bicrystal grain boundary model of Σ3 GB [51] showing XZ cross-sectional structure. The black dashed line represents the grain boundary plane separating the grain I (GI) and grain II (GII). Labels with I, Si, and C represent interstitial and substitutional defect sites.

4. **Results & Discussion**

Despite many experimental studies on the release of Cs from TRISO particles and the SiC layer alone, there exists an ambiguity in the quantitative analysis of Cs solubility and how the diffusion of Cs occurs through the SiC layer. Shrader *et al*. [9] have reported the theoretical studies on the diffusion of Cs through bulk cubic 3C-SiC using density functional theory and proposed that the Cs transport in bulk SiC follows a "ring mechanism", which uses the most stable defect cluster of ($Cs_C$−$2V_{Si}^{3-}$) as a template for the movement of Cs from one carbon site to the neighboring carbon site via the formation of other less-stable mixed vacancy clusters. The energy barrier for Cs to move from one carbon site to another using the two Si vacancies was found to be 5.14 eV, and this number compared well with the 5.21 eV found by Minato *et al*. [26] in the high-temperature regime using integral release experiments. Though the theoretical activation energy compared well with that from the experiments, the high formation energies (from 10 – 28 eV) for various types of Cs substitutional defects and the relevant charge states in bulk-SiC suggested a vanishing solubility of Cs in bulk SiC [9]; even the lowest energy defect ($Cs_C-2V_{Si}^{3-}$) involved in the "ring-mechanism" was found to have a formation energy of 10.3 eV. However, the integral release experiments on Cs [3, 11, 52] suggest a reasonable solubility of Cs in SiC based on their data on activation energies at low temperatures. Prompted by these findings, Shrader *et. al.* [9] proposed further investigation of Cs in SiC.

In the present work, we investigate the defect formation energies of Cs at the Σ3 grain boundary of SiC to investigate the Cs diffusion at the grain boundaries and also to address the question of Cs solubility in SiC. In line with the integral release experiments performed at low temperatures [3, 11, 52], we find that Cs defect formation energies at the GBs (0.2 – 9.5 eV) are much lower than those (10.8 – 28.6 eV) found for Cs in bulk-SiC. We believe that these low defect formation energies could translate into low activation energy barriers observed in the experiments [3, 11, 52] for Cs diffusion at low temperatures. Low formation energies at GBs also resolve the problem of vanishing solubility of Cs in SiC. We have also compared our results of Cs at the Σ3 GB with those of Cs in the bulk SiC reported previously by Shrader *et al*. [9] to analyze the effect of grain boundaries in the release of Cs. Further, we have compared our observations for Cs at Σ3 GBs with those of other fission products like Ag [21] at the grain boundaries to gain understanding on the role of the chemical nature and size of the fission products in their diffusion through SiC.

## 4.1 Defects in Bulk 3C-SiC

Since the diffusion processes are known to be mediated through intrinsic defects such as vacancies, our investigation started with evaluating the defect formation energies of the intrinsic defects in bulk SiC using a 2×2×2 super cell. The defect formation energies for Si and C vacancies in bulk SiC are shown in Table 1. Note that these numbers agree well with those reported by Sharder *et al.* [9, 21] and with data from other literature [53-56]. The defect formation energies showed the formation of the carbon vacancy to be relatively easier compared to that of the silicon vacancy.

**Table 1** Intrinsic defect formation energies (DFE, eV) in bulk 3C-SiC and their comparison with the values reported in literature.

| Defect | Defect formation energies (eV) | | | | | |
|---|---|---|---|---|---|---|
| | Present work | [9, 21] | [53] | [54] | [55] | [56] |
| $V_{Si}$ | 7.62 | 7.63 | 7.62 | 6.64 | 7.48 | 7.01 |
| $V_C$ | 4.38 | 4.13 | 3.47 | 5.48 | 3.63 | 4.89 |

## 4.2 Defects at Σ3-Grain Boundary of SiC

The integral release experiments on the Cs diffusion through SiC show a wide range of diffusion coefficients magnitudes associated with a range of activation energies in distinct temperatures zones [3, 8, 11, 12, 25, 26, 52], suggesting a temperature dependence of the Cs diffusion through the SiC layer. This substantial change in the activation energy as a function of temperature indicates that bulk diffusion dominates at high temperatures [9], whereas GB diffusion could prevail at low temperatures, which has not been confirmed yet. Therefore, we look at the defect formation of Cs at the Σ3 grain boundaries, which is a special low- energy twin boundary found versatile in cubic 3C-SiC[37]. We have investigated Cs interstitial and substitutional defects, along with Cs-vacancy defect clusters with up to two NN vacancies.

### 4.2.1 Intrinsic Defects at Grain Boundary of SiC

The defect formation energies for a few intrinsic defects at the Σ3 grain boundary, in the GI and GII (shown in Figure 1) were calculated in the 2×2×1 supercell having 256 atoms. These intrinsic defect formation energies for various types of defects are reported in Table 2, where Si-rich conditions were used for the Si and C chemical potentials (Eq. 1). Table 2 also shows the comparison of these numbers with the previously reported values [20]. In agreement with the previous calculations [9, 20, 21], we find that the formation of the C vacancies is energetically more favorable than the formation of Si vacancies. Note that a similar trend was observed for bulk-SiC. The formation energies for C and Si vacancies in GI and in GII are found to be different, which can be attributed to the variation in the atomic structures of the two grains. Overall, our numbers are found to be in excellent agreement with the numbers reported previously [20], where the VASP DFT simulation package was used. This comparison gives us confidence in using our pseudopotentials to study Cs at the Σ3 grain boundary of SiC. It also shows that the GB structure is the same as that used in previous studies [20].

**Table 2** Defect formation energies (DFE, eV) for intrinsic defects in the bulk and at the grain boundary of SiC. Numbers in the parentheses are obtained from the reported literatures [20, 21]

|             | Site      | DFE (eV)    |
| ----------- | --------- | ----------- |
| Bulk SiC    | $V_C$     | 4.38 (4.13) |
|             | $V_{Si}$  | 7.62 (7.63) |
| Grain Plane | C1        | 3.08 (2.78) |
|             | Si1       | 4.89 (4.13) |
| Grain I     | C2        | 2.20 (1.97) |
|             | Si3       | 5.71 (5.81) |
| Grain II    | C2        | 2.18 (1.32) |
|             | Si3       | 5.26 (4.52) |

We find that the defect formation energies of C and Si vacancies in bulk SiC [DFE($V_C$) = 4.38 eV and DFE($V_{Si}$) = 7.62 eV] are higher than those in the Σ3-grain boundary [DFE($V_C$) = 2.18 eV and DFE($V_{Si}$) = 4.89 eV], indicating an energetically unfavorable Cs substitution in bulk compared to that in the Σ3-grain boundary. In general, the defect formation energies for the intrinsic defects in the grain boundary are lower than those of the bulk 3C-SiC. This observation may be attributed to the higher energy requirement for the destabilization of the symmetric atomic structure in the case of bulk SiC, whereas the formation of similar defects in the grain boundary between GI and GII are not that energy demanding due to the presence of disorder in the structure.

**4.2.2   Cs Point Defects in the Grain Boundary of SiC**

Since the propagation of Cs occurs initially by its incorporation at the vacant sites, we have calculated the defect formation energies of Cs as isolated point defects in the grain boundary of SiC. Cs was incorporated at all types of carbon and silicon vacancy sites in GI, GII, and GBs (shown in Figure 1), and our results on the defect formation energies are shown in Table 3. In Figure 2, we plot the range of formation energies for various defects in grain boundaries by solid black lines and the corresponding lowest formation energies in bulk by black (red) dashed lines for neutral (charged) defects. Figure 2 also shows the comparison with Ag defects in bulk and grain boundaries. Lowest formation energies for Cs in bulk SiC are found to be in the charge neutral state in agreement with Ref. [9]. For Ag, however, the lowest formation energy in bulk does not necessarily correspond to the neutral Ag defects. For example, $Ag_{Si}$ has the lowest formation energy in the charge state of -3. Notice that the lowest formation energies of $Cs_C$ and $Cs_{Si}$ in the GB are lower than those of $Ag_C$ and $Ag_{Si}$ in spite of the bigger size of Cs than Ag. Further, the overall differences between the highest defect formation energy in grain boundary and the lowest DFE in bulk SiC are found to be much greater for Cs than Ag.

These large differences between lowest bulk and highest GB formation energy will have a significant impact on the segregation factor for Cs in GBs as it is defined as $s = C_{GB}/C_{bulk}$, where $C_{GB}$ and $C_{Bulk}$ are the defect concentrations in GBs and in bulk, respectively. In general, the defect concentration is defined as

$$C^i = N^i \times \exp\left(-\frac{E_f^i}{kT}\right), \qquad (2)$$

where, $N^i$ is the number of sites of type $i$ in bulk or GB and $E_f$ is their formation energy. Using this equation, the segregation factor can be rewritten as

$$s = \frac{N_{GB}}{N_{Bulk}} \exp\left(\frac{E_{F,Bulk} - E_{F,GB}}{kT}\right) \qquad (3)$$

Thus, the segregation factor has an exponential dependence on the formation energy difference between bulk and GB. Since there is a range of formation energies for a given defect in GB, we use the highest formation energy in GB and the lowest formation energy in bulk to calculate the energy difference in equation (3). Reason for choosing highest and lowest DFEs in GB and bulk is the following: $N_{GB}$ in GB would be much smaller than the $N_{Bulk}$ as the GB volume to bulk volume ratio in SiC is about 0.0019 [20]. This means that there are a much fewer ($\sim 10^{-3}$) defect sites available in GB than in bulk for a given defect type. As soon as lower energy defect sites get occupied, next higher energy defect sites will get occupied and so on. Therefore, almost the whole range (shown by two black solid lines in Figure 2) of defect sites may get occupied as more and more FPs are being released from the kernel. Since there is a large bulk volume available, it is mostly the lowest energy defect site that will be occupied at low to medium temperatures. It is in this sense that we use lowest formation energy for defects in bulk and the highest formation energy in grain boundaries. From Figure 2, note that this energy difference is much greater for Cs than for Ag for all the types of isolated defects. For $Ag_{Si}$, the lowest formation energy in bulk lies below the lowest formation energy in GB.

Specifically, the overall difference between the highest defect formation energy in the grain boundary and the lowest in bulk SiC for the $Cs_C$ and $Ag_C$ are ~7.0 eV and ~1.8 eV, respectively. Similarly, overall difference between the highest defect formation energy in the grain boundary and the lowest in bulk SiC for $Cs_{Si}$ and $Ag_{Si}$ is found to be ~7.8 eV and ~-0.1 eV, respectively. These differences for interstitial defect sites are even greater, ~17.0 eV and ~2.8 eV for $Cs_{int}$ and $Ag_{int}$, respectively. All these data suggest that Cs has a much greater probability of getting incorporated at the grain boundary than in bulk, whereas Ag has a more or less equal probability of getting incorporated in bulk and at the grain boundaries, especially through the formation of $Ag_{Si}$ and $Ag_C$ defects. Low defect formation energies of Cs at the Σ3 grain boundary of SiC also suggest a reasonable solubility of Cs in SiC at low temperatures as found in experiments [3, 11, 52]. We note that Cs solubility in SiC was also in question from previous theoretical calculations [9] in bulk SiC.

While observing the individual defect formation energies of Cs at various sites, we find that Cs has the lowest formation energy of 2.31 eV at the silicon site (Si4, Figure 1) present in the GII and the highest defect formation energy of 9.52 eV at an interstitial site (I1, Figure 1) made of a 5-membered ring present in the grain boundary. Whereas, for Ag, the lowest defect formation energy was found to be 3.63 eV at the carbon site (C2, Figure 1) in GI, and the highest was found to be 7.65 eV at an interstitial site (I4, Figure 1) made of a 6-membered ring in GII.

Table 3 Defect formation energies (eV) for Cs substitution at carbon, silicon and interstitial sites in bulk, grain plane (GB), grain-I(GI), and in grain-II(GII) of the Σ3 grain boundary of SiC. Note the values on the parentheses are from the Ref. [9]

| | C-Site | DFE(Cs) (eV) | DFE(Ag) (eV) [20] | Si-Site | DFE(Cs) (eV) | DFE(Ag) (eV) [20] | I-Site | DFE(Cs) (eV) | Int-Site | DFE(Ag) (eV) [20] |
|---|---|---|---|---|---|---|---|---|---|---|
| Bulk SiC | Bulk | 12.07 (*12.50*) | 7.39 | Bulk | 14.58 (*12.71*) | 6.60 | Bulk | 26.55 (*23.46*) | Bulk | 10.49 |
| Grain Plane | C1 | 4.72 | 5.60 | Si1 | 2.74 | 4.01 | I1 | 9.52 | I | Unstable |
| | | | | | | | I2 | 5.88 | II | 6.71 |
| | | | | | | | | | IIa | 4.70 |
| GI | C2 | 3.70 | 3.63 | Si2 | 4.69 | 4.47 | I3 | 5.96 | | |
| | C3 | 2.54 | 4.59 | Si3 | 6.76 | 6.15 | I4 | Unstable | | |
| | C4 | 4.61 | 4.99 | Si4 | 3.38 | 3.95 | | | | |
| G II | C2 | 5.00 | 4.22 | Si2 | 4.20 | 4.75 | I3 | 4.92 | IIIL | 5.99 |
| | | | | | | | | | IVL | 7.65 |
| | C3 | 4.74 | 4.76 | Si3 | 5.80 | 5.18 | I4 | 7.57 | IIILa | 3.27 |
| | C4 | 4.26 | 4.84 | Si4 | 2.31 | 3.69 | | | IVLa | 7.38 |

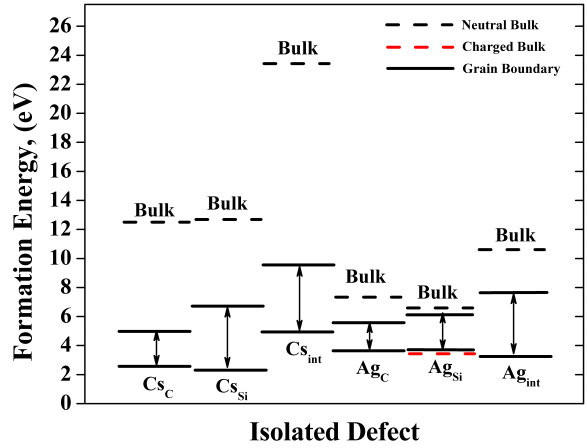

**Figure 2** Formation energy ranges for isolated Cs defects in Σ3 grain boundaries (solid black lines) and their comparison with Ag defects. Dashed horizontal lines represent the corresponding lowest formation energies in bulk SiC. Black (red) dashed lines correspond to neutral and charged defects in bulk.

### 4.2.3 Cs – Single Vacancy Defects Cluster at Grain Boundary of SiC

It is known that the movement of fission products becomes more favorable through the formation of defect clusters where the fission products are bound to a single or multiple vacancies [9, 20, 21]. In the present study, we have calculated the defect formation energies of the defect clusters having Cs substitution at either the silicon or carbon site along with one or two nearest neighbor silicon or carbon vacancies. Our results on the defect formation energies of Cs occupying the carbon or silicon site along with a single vacancy of either silicon or carbon, respectively, are shown in **Table 4**. In each case, there are a total of 4 nearest neighbors (NN) of opposite type, i.e., when Cs occupies a carbon site, four types of Si vacancies could form and four types of C vacancies could form when Cs occupies the Si site.

Formation energy ranges are plotted in Figure 3 along with those of Ag and the bulk data. Notice that the lowest DFEs of Cs–single vacancy clusters are also lower in energy than those of Ag-single vacancy clusters as in the case of isolated defects. We find that the defect formation energies of $Cs_{Si}+V_C$ are in the range of 0.23 – 4.99 eV; however its formation energy in bulk SiC is 11.97 eV. On the other hand, the formation energy of $Ag_{Si}+V_C$ in the grain boundary was found to be in the range of 2.76 eV – 4.60 eV [20], while the corresponding DFE in bulk was found to be 5.32 eV [21]. Further, we find that the differences between the highest DFEs in the grain boundary and the lowest in bulk SiC for $Cs_{Si}+V_C$ and $Ag_{Si}+V_C$ are ~8.0 eV and ~0.7 eV, respectively. Similarly, the overall difference between the highest DFE in the grain boundary and the lowest in bulk SiC for $Cs_C+V_{Si}$ and $Ag_C+V_{Si}$ are 7.75 eV and 0.72 eV, respectively. This significant difference in the DFEs of $Cs_{Si}+V_C$ between the bulk and the grain boundary suggests that Cs will completely populate GBs before getting dissolved in bulk. Since the formation energies of Cs at GBs are so much lower than those in bulk, the activation energies for Cs diffusion at GBs could also be lower than those in bulk. Activation energies in GBs are generally found to be lower than those in bulk (refs). Therefore, low activation energies of 1.3 – 2.5 eV are observed for Cs diffusion in the temperature range of 1073 – 1873 K and the activation energy jumps to 5.3 eV in the 1873 – 2173 K temperature range. This suggests that GB diffusion will be dominant at lower temperatures and bulk diffusion at higher temperatures. Whereas, the formation energies of Ag at GBs do not decrease as much, therefore Ag will start occupying GBs at lower temperatures and as the temperature slightly increases, it will quickly start to get dissolved in the bulk. Hence, activation energies of Ag are found to be 1.83 – 2.5 eV throughout the temperature range used in experiments. We also note that the overall formation energies of the $Cs_{Si/C}+V_{C/Si}$ are lower than those of $Ag_{Si/C}+V_{C/Si}$.

While observing the individual defect formation energies of single vacancy defect clusters involving $Cs_{Si/C}$ point defects in the grain boundary of SiC, we find that the lowest defect formation energy for both cluster types ($Cs_C+V_{Si}$, $Cs_{Si}+V_C$) to be in the GII, whereas the highest defect formation energies for $Cs_C+V_{Si}$ and $Cs_{Si}+V_C$ are found to be in GI and GII, respectively. Note that the overall defect formation energies of the single vacancy defect clusters have further decreased by about 1.5 eV compared to those of the isolated $Cs_{Si/C}$ defects at the grain boundary, $Cs_{Si}+V_C$ is found to have a formation energy of only 0.23 eV. The significant difference in the defect formation energies for the Cs-single vacancy defect clusters in grain boundary and in bulk-SiC further increases the likelihood of Cs incorporation in grain boundaries than in bulk.

We have performed Nudged Elastic Band (NEB) calculations to find activation energies for Si and C vacancy hopping with Cs substituted at C and Si sites, respectively. For Cs incorporated at the C4 site, we find an activation energy of 3.70 eV for Si vacancy hopping from Si3 to Si4. In the reverse direction, this barrier is found to be 4.77 eV. For Cs incorporation at Si4 site, the activation energy for carbon vacancy hopping from C4 to C3 is found to be 6.74 eV in the forward direction and 5.88 eV in the reverse direction.

**Table 4** Defect formation energies of $Cs_{C/Si}$-$V_{Si/C}$ types of defect cluster in grain I (GI) & grain II (GII) of Σ3-grain boundary of SiC.

| Site | Vacancy | DFE (eV) | Site | Vacancy | DFE (eV) |
|---|---|---|---|---|---|
| GII-C2 | $V_{Si2}$ | 1.76 | GI-Si2 | $V_{C2}$ | 3.06 |
| | $V_{Si3}$ | 4.88 | | $V_{C3}$ | 2.01 |
| GI-C3 | $V_{Si4}$ | 2.16 | GII-Si2 | $V_{C2'}$ | 3.04 |
| | $V_{Si1}$ | 1.67 | | $V_{C2}$ | 1.75 |
| | $V_{Si2}$ | 2.01 | | $V_{C3}$ | 3.96 |
| | $V_{Si4'}$ | 2.04 | | $V_{C3'}$ | 3.96 |
| GII-C3 | $V_{Si2}$ | 4.18 | GII-Si3 | $V_{C1}$ | 4.28 |
| | $V_{Si4}$ | 1.82 | | $V_{C2}$ | 4.99 |
| | $V_{Si1}$ | 1.56 | | $V_{C4}$ | 2.92 |
| GI-C4 | $V_{Si2'}$ | 3.10 | GI-Si4 | $V_{C4}$ | 2.63 |
| | $V_{Si3}$ | 2.94 | | $V_{C3}$ | 2.73 |
| | $V_{Si4}$ | 2.96 | | $V_{C3'}$ | 2.74 |
| GII-C4 | $V_{Si4}$ | 1.89 | GII-Si4 | $V_{C3}$ | 1.82 |
| | $V_{Si3}$ | 2.95 | | $V_{C4}$ | 0.96 |

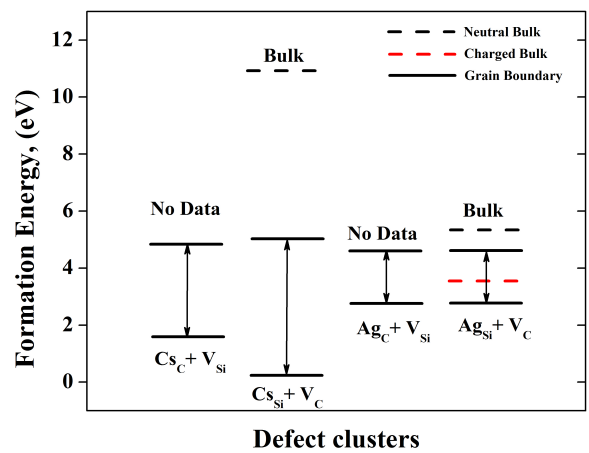

**Figure 3.** Formation energy ranges for Cs-Single vacancy defect clusters in Σ3 grain boundaries (solid black lines) and their comparison with Ag-single vacancy defect clusters. Dashed horizontal lines represent the corresponding lowest formation energies in bulk SiC. Black (red) dashed lines correspond to neutral (charged) defects in bulk.

**4.2.4 Cs-Divacancy Defect Clusters at Grain Boundary of SiC**

Transport of fission products through a material can be further facilitated by the increase in the intrinsic defect concentration [9]. Therefore, in the present work Cs-divacancy defect clusters are also studied at the grain boundary of SiC. When Cs occupies either silicon or carbon sites, there can be 6 combinations of carbon or silicon divacancies, respectively. The ranges of defect formation energies for various combinations of vacancies along with their bulk counterparts are shown in Figure 4. For comparison, we also plot the formation energy ranges of Ag-divacancy clusters from Ref.[20] and the corresponding numbers in the bulk SiC [21] in this figure. The formation energies of $Cs_C+2V_{Si}$ and $Cs_{Si}+2V_C$ defect clusters are found to be in the range of 2.20 eV – 3.99 eV and 1.04 eV – 2.27 eV, respectively. Consequently, some of the single vacancy defect clusters are lower in energy than divacancy defect clusters. This may be due to the energy cost of creating an extra intrinsic defect. However, in the case of $Cs_{Si}+V_C$, many of the single vacancy clusters have higher energies than the divacancy clusters suggesting that Cs prefers to have two carbon vacancies when incorporated at Si-site. In general, single vacancy defect clusters are found to have a broader range of formation energies than divacancy defect clusters. By contrast, the formation energies of $Ag_C+V_{Si}$ were found to be distinctively lower than those of divacancy clusters [20, 21]. Formation energies of $Ag_{Si}+V_C$ were also lower than those of divacancy clusters but with slight overlap between the two ranges [20, 21]. Cs-divacancy defect clusters in bulk were also found to have higher formation energy than the single-vacancy defect clusters [9].

Further, the formation energies of $Cs_{Si}+2V_C$ and $Cs_C+2V_{Si}$ in the GBs are much lower than those of in bulk which were found to be 12.37 eV and 13.87 eV respectively. On the other hand, $Ag_{Si}+2V_C$ defect cluster in the grain boundary was found to have energies between 3.74 eV and 5.73 eV [20], while $Ag_{Si}+2V_C$ in bulk has the formation energy 6.44 eV [21]. It can be noted again that the formation energies of $Cs_{Si}+2V_C$ and $Cs_C+2V_{Si}$ are lower than those of $Ag_{Si}+2V_C$ and $Ag_C+2V_{Si}$, respectively. As a result, we find that the overall difference between the highest defect formation energies in the grain boundary and in bulk SiC for the $Cs_{Si}+2V_C$ and $Ag_{Si}+2V_C$ are 10.1 eV and 0.71 eV, respectively. Similarly, the overall difference between the highest defect formation energies in the grain boundary and in bulk SiC for $Cs_C+2V_{Si}$ and $Ag_C+2V_{Si}$ is 9.88 eV and 3.1 eV, respectively. All these data for the defect formation energies of divacancy defect clusters further suggest a greater probability of Cs getting incorporated at the grain boundary than in bulk SiC.

While observing the individual divacancy clusters at the grain boundary, the lowest formation energy for $Cs_C+2V_{Si}$ is found to be at the GI, while that of $Cs_{Si}+2V_C$ is at the GII; whereas the highest defect formation energy for both types of clusters is found in the GI. Overall, the stabilization of isolated Cs point defects and defect clusters in the grain boundary of SiC shows Cs will get implanted in the grain boundary more than in bulk SiC. Stabilization of Cs-vacancy clusters over isolated defects shows that Cs movement through the grain boundaries will occur via the formation of Cs-vacancy clusters. As observed in bulk SiC [9], Cs does not exhibit a site preference for defect formation at the grain boundaries as well.

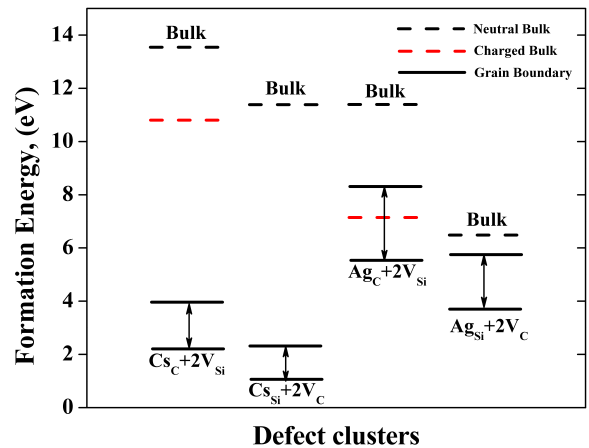

**Figure 4**. Defect formation energy ranges of $Cs_{C/Si}+2V_{Si/C}$ defect clusters in the grain I (GI) and grain II (GII) of the Σ3 grain boundary of SiC. The dashed horizontal lines represent the corresponding bulk values.

**4.3**      Comparison of Cs with Ag in SiC

In Table 5, we show reported values of diffusion coefficients and activation energies of Cs and Ag in various temperature ranges obtained from integral release and ion implantation experiments. The top (bottom) part of the table (thick bordered region) shows the data for Cs (Ag). Notice that Cs exhibits a very clear bimodal diffusion behavior, i.e., its diffusion coefficients change by at least seven orders of magnitude and the activation energy changes by ~3 eV while going from the low to high temperatures. Top part of the table is divided in unshaded and shaded regions representing these two modes of diffusion of Cs in SiC. However, no such demarcation is observed for Ag at low and high temperatures. We believe that the presence of such a demarcation for Cs is a result of the large energy differences between GB and bulk formation energies. The absence of such a demarcation for Ag is due to the very small or non-existent energy differences between GBs and bulk. In some cases, formation energies of Ag in bulk are even lower than those in GBs.

**Table 5** Diffusion coefficients and activation energies for Cs and Ag in SiC obtained from various experiments reported in literature.

| Reference | $D_0$ (m$^2$/s) | Q (eV/atom) | Temp (K) |
|---|---|---|---|
| Cs, [3] | $5.5 \times 10^{-14}$ | 1.30 | 1073-1673 |
| Cs, [11] | $3.5 \times 10^{-09}$ | 2.45 | 1273-1773 |
| Cs, [52] | $1.77 \times 10^{-11}$ | 1.82 | 1273-1873 |
| Cs, [52, 57] | $2.50 \times 10^{-02}$ | 5.21 | 1873-2173 |
| Cs, [3] | $1.6 \times 10^{-02}$ | 5.33 | 1673-1973 |
| Ag, [58] | $3.60 \times 10^{-9}$ | 2.23 | not given |
| Ag, [59] | $6.76 \times 10^{-9}$ | 2.21 | 1073-1773 |
| Ag, [60] | $5.00 \times 10^{-10}$ | 1.89 | 1273-1773 |
| Ag, [60] | $3.50 \times 10^{-10}$ | 2.21 | 1473-2573 |
| Ag, [60] | $3.60 \times 10^{-9}$ | 2.23 | 1273-1773 |
| Ag, [60] | $6.80 \times 10^{-11}$ | 1.83 | 1473-1673 |
| Ag, [11] | $4.50 \times 10^{-9}$ | 2.26 | 1273-1773 |
| Ag, [27] | $4.30 \times 10^{-12}$ | 2.50 | 1473-1673 |

## 5. Conclusions

By calculating the formation energies of substitutional Cs point defects and defect clusters at the Σ3 grain boundaries, we show that Cs is much more stable at the grain boundaries of SiC than in the bulk. We note that the differences between the highest defect formation energies at the grain boundary and the lowest in bulk SiC for all types of defects (point defects, the single vacancy cluster and divacancy cluster) involving substitutional Cs are in the range of 7.07 eV and 17.03 eV. This suggests that Cs will predominantly get incorporated in $\Sigma 3$ grain boundaries at low temperatures and/or it has a strong tendency to segregate to the GBs. Low defect formation energies of Cs will also result in lower activation energies in the grain boundaries. In experiments, the activation energies are found to be between 1.3 and 2.5 eV at temperatures below 1400 $^o$C, whereas, it jumps to ~5 eV above this temperature. This large variation in the activation energies at low and high temperatures correlates well with the large difference in formation energies between grain boundaries and bulk and confirms the hypothesis proposed in Ref. [9] that grain boundary diffusion dominates at low temperatures and bulk diffusion at high temperatures. By contrast, the defect formation energy differences for Ag were found to be in the range of 0.45 eV – 3.10 eV [20, 21], which suggests that Ag gets uniformly distributed through grain boundaries and bulk of SiC at intermediate to high temperatures and therefore Ag does not exhibit the change in dominant diffusion mechanism as a function of temperature as observed in experiments [11, 14, 60]. Another interesting result we find is that the defect formation energies of Cs are lower than those of Ag in $\Sigma 3$ grain boundaries, while Cs has much higher formation energies than Ag in bulk SiC. Thus, Cs and Ag have a very different behavior in SiC.

The large versus small formation energy differences of FPs in grain boundaries and bulk can have far-reaching consequences on controlling the release of FPs through the structural component in TRISO fuel particles. For example, large energy differences for Cs suggest that Cs release from SiC can be controlled via grain-boundary-engineering by reducing the percentage of $\Sigma 3$ grain boundaries in the cubic SiC. Tsurekawa *et al*. [61] showed that the percentage of $\Sigma 3$ grain boundaries in SiC can be reduced by doping SiC with Mg and Phosphorus. However, one has to first study the FP-dopant interactions in SiC and confirm that FPs do not form complexes with the dopant species which can facilitate the diffusion of FPs through SiC. Σ3 grain boundaries can also be reduced by controlling the coating rate of SiC on the TRISO particle such that a uniform particle distribution of SiC is obtained which is also shown to decrease the percentage of Σ3 grain boundaries [37, 62]. On the other hand, the marginal differences in the defect formation energies of Ag at grain boundary and in bulk of SiC pose a challenge to control Ag release through SiC [20, 21].

**Acknowledgements**

We wish to thank Izabela Szlufarska and Dane Morgan from the University of Wisconsin at Madison for helpful discussions and for providing the grain boundary structure. R.P. gratefully acknowledges the financial support from the Research Corporation Cottrell college science award. Calculations were performed at the HPC center of Idaho National Laboratory.